\documentclass[12pt]{article}

\pdfoutput=1

\usepackage{graphics}
\usepackage{epsfig}
\usepackage{amsfonts}
\usepackage{amssymb}
\usepackage{amsmath}
\usepackage{dsfont}
\usepackage{pifont}
\usepackage{bbm}
\usepackage{multirow}
\usepackage{latexsym}
\usepackage{verbatim}
\usepackage{mcite}
\usepackage{cite}
\usepackage{units}
\usepackage[all]{xy}
\usepackage{cancel}
\usepackage{slashed}

\def\a{{\alpha}}
\def\b{{\beta}}

\def\bfone{\relax{\rm 1\kern-.35em 1}}

\newcommand{\cN}{{\cal N}}

\newcommand{\be}{\begin{equation}}
\newcommand{\ee}{\end{equation}}
\newcommand{\ben}{\begin{displaymath}}
\newcommand{\een}{\end{displaymath}}
\newcommand{\bea}{\begin{eqnarray}}
\newcommand{\eea}{\end{eqnarray}}

\newcommand{\bean}{\begin{eqnarray*}}
\newcommand{\eean}{\end{eqnarray*}}

\DeclareMathAlphabet{\mathpzc}{OT1}{pzc}{m}{it}

\begin{document}
\pagestyle{plain}


\makeatletter \@addtoreset{equation}{section} \makeatother
\renewcommand{\thesection}{\arabic{section}}
\renewcommand{\theequation}{\thesection.\arabic{equation}}
\renewcommand{\thefootnote}{\arabic{footnote}}


\setcounter{page}{1} \setcounter{footnote}{0}


\begin{titlepage}

\begin{flushright}
UUITP-24/13\\
\end{flushright}

\bigskip

\begin{center}

\vskip 0cm

{\LARGE \bf An alternative to anti-branes and O-planes?} \\[6mm]

\vskip 0.5cm

{\bf Ulf Danielsson \,and\, Giuseppe Dibitetto}\\

\vskip 25pt

{\em Institutionen f\"or fysik och astronomi, \\ 
University of Uppsala, \\ 
Box 803, SE-751 08 Uppsala, Sweden \\
{\small {\tt \{ulf.danielsson, giuseppe.dibitetto\}@physics.uu.se}}} \\

\vskip 0.8cm

\end{center}

\vskip 1cm

\begin{center}

{\bf ABSTRACT}\\[3ex]

\begin{minipage}{13cm}
\small

In this paper we consider type IIA compactifications in the isotropic $\mathbb{Z}_{2}\,\times\,\mathbb{Z}_{2}$ orbifold with a flux-induced perturbative superpotential combined with non-perturbative effects.
Without requiring the presence of O-planes, and simply having D$6$-branes as local sources, we demonstrate the existence
of de Sitter (dS) critical points, where the non-perturbative contributions to the cosmological constant have negligible size. We note, however, that these solutions generically have tachyons.  

By means of a more systematic search, we are able to find two examples of stable dS vacua with no need for anti-branes or O-planes, which, however, exhibit important non-perturbative corrections. 
The examples that we present turn out to remain stable even after opening up the fourteen non-isotropic moduli.  

\end{minipage}

\end{center}

\vfill

\end{titlepage}


\tableofcontents

\section{Introduction and Motivation}
\label{sec:introduction}

The problem of constructing metastable de Sitter solutions in string theory is of paramount importance. A rich class of proposed solutions were provided by KKLT for type IIB string theory \cite{Kachru:2003aw}.
There are three important ingredients that are crucial for these constructions: anti-branes, O-planes, and non-perturbative corrections. Each of these have their own particular problems associated with them. 
The anti-branes break suspersymmetry explicitly, and it has been claimed (see \emph{e.g.} refs~\cite{Bena:2009xk}, and references therein) that they introduce harmful divergences that 
signal the presence of dangerous instabilities. The O-planes are ill-understood structures to be viewed as non-dynamical, and for which no description in terms of dynamical degrees of freedom is known.  
Finally, the non-perturbative corrections, which may have various origins in this context (\emph{e.g.} brane instantons \cite{Uranga:2008nh}, or gaugino condensation \cite{Barreiro:1998nd}), 
are known only in their general form and very little is known about their exact dependence on all moduli fields.

In ref.~\cite{Saltman:2004sn} an attempt was made to get rid of the anti-branes. The equations of motion were solved in all directions except the volume modulus, which was kept constant by hand. 
Solutions were found without tachyons where all complex structure moduli were massive, and a few K\"ahler moduli were massless. It was then claimed that the addition of non-perturbative corrections could 
stabilise all the K\"ahler directions, including the run-away. As a result, ref.~\cite{Saltman:2004sn} claimed to provide a mechanism for producing metastable de Sitter solutions without the need of 
anti-branes. 

In this paper we take one further step, focusing on type IIA, and omit also the O-planes. Contrary to ref.~\cite{Saltman:2004sn}, we explicitly check for stability also after the proper introduction of 
the non-perturbative terms. The reason to expect that you do not necessarily need O-planes in such a set-up is to be found in ref.~\cite{Blaback:2013fca}, where the proof that dS solutions obtained by means of perturbative fluxes need O-planes was extended to ``quasi-dS'' solutions, \emph{i.e.} whenever $V\,>\,0$, and 
$\epsilon_{V}\,\equiv\,\frac{1}{2}\,\left(\frac{\partial_{I}V}{V}\right)^{2}\,<\,1$. The argument was based on writing $V$ in terms of the eom of the universal modulus $\tau$
\be
\label{dS_O}
V \, = \, V_{H_{3}} \, + \, V_{\omega} \, + \, \sum\limits_{p}{V_{F_{p}}} \, + \, V_{\textrm{O}6/\textrm{D}6} \, = \, 
 - \, \frac{1}{2} \, \tau \, \partial_{\tau}V \, \underbrace{- \, \sum\limits_{p}{f_{p}^{2} \, \rho^{3-p} \, \tau^{-4}}}_{\leq \,\, 0} \, - \, \frac{1}{2} \, N_{6} \, \tau^{-3} \ ,
\ee
where $V \, \overset{!}{>} \, 0$ implies $N_{6} \, < \, 0$ if the first term on the rhs is zero (dS) or not large enough (quasi-dS).

Following the philosophy of ref.~\cite{Saltman:2004sn}, one can now consider the possibility of lifting the $\tau$ eom at a perturbative level, hence leaving it as a run-away direction, and hope for some 
non-perturbative effects to fix it. One can easily get convinced that, by doing so, it is indeed possible to find examples of backgrounds with perturbative fluxes stabilising all directions but $\tau$ 
and $N_{6}\,>\,0$, \emph{i.e.} pure D-brane configurations without an orientifold.

Moreover, in the type IIA duality frame with O$6$/D$6$, one expects a class of non-perturbative effects associated with open-string dynamics living on the D$6$-branes to introduce an exponential
dependence in $\tau$. In order to see this, we need to observe that the YM coupling for the gauge theory living on the branes scales as the volume of the corresponding wrapped cycle \cite{Kachru:2003aw}.
Focusing purely on universal scalars (see \eqref{non-pert-S} \& \eqref{non-pert-T} with $\sigma=1$), each non-perturbative contribution to the superpotential scales as $e^{-1/g_{\textrm{YM}}^{2}}$ where
\be
\begin{array}{lclclclc}
\dfrac{1}{g_{\textrm{YM}}^{2}} & = & \dfrac{\textrm{vol}_{3}}{g_{s}} & \sim & \dfrac{\rho^{3/2}}{g_{s}} & \equiv & \tau & .
\end{array}
\ee

Keeping this as a general motivation, the aim of this paper is two-fold. Firstly, we will show that, given a flux background admitting a solution where all the moduli but one (say $\phi$) are stabilised, it is possible
to stop the run-away of $\phi$ by adding a non-perturbative effect involving it. We will see that this can be done without introducing neither anti-branes nor O-planes, and within a regime where the non-perturbative 
contributions to the cosmological constant and all the other eom's are negligible, whereas $\partial_{\phi}V$ receives important non-perturbative contributions, as it should in order for them to stabilise
 $\phi$. 
However, we will also see that this generically leads to big off-diagonal mixing in the mass matrix between $\phi$ and all the other scalars, which generically yields tachyons. However, this might still leave room for fine-tuned critical points where this undesired feature does not show up.  

Secondly, by carrying out a more systematic search based on a technique introduced in ref.~\cite{Danielsson:2012by}, we are able to find fully stable de Sitter solutions with neither anti-branes nor 
orientifolds, even though their corresponding non-perturbative contributions to the energy are important.  
Still, the solutions we find have many attractive features, and we believe that they deserve further study. In particular, we discuss the separation of various scales such as the Hubble scale, the size 
of the extra dimensions, the string scale, and the Planck scale.

\textbf{Note added:} Upon completion of this manuscript we became aware of \cite{Blaback:2013qza}, where a similar approach is used to examine the case of type IIB. In this work, one also allows for an explicit dependence in the prefactors appearing inside the non-perturbative superpotential on the complex structure moduli.

\section{Type IIA with fluxes and non-perturbative effects}
\label{sec:review}

We will focus here on a class of theories arising from isotropic $T^{6}/(\mathbb{Z}_{2}\,\times\,\mathbb{Z}_{2})$ orbifold compactifications of type IIA with D$6$-branes. These string compactifications 
have particular $\cN=1$ supergravities known as $STU$-models as low-energy effective descriptions. These enjoy $\textrm{SL}(2)^{3}$ global symmetry and contain three complex scalars 
$\Phi^{\a}\,\equiv\,(S,T,U)$ spanning $\left(\textrm{SL}(2)/\textrm{SO}(2)\right)^{3}$.

The kinetic Lagrangian can be derived from the following K\"ahler potential
\be
\label{Kaehler_STU}
K\,=\,-\log\left(-i\,(S-\overline{S})\right)\,-\,3\,\log\left(-i\,(T-\overline{T})\right)\,-\,3\,\log\left(-i\,(U-\overline{U})\right)\ .
\ee
This yields
\be
\mathcal{L}_{\textrm{kin}} = 
\frac{\partial S\partial \overline{S}}{\left(-i(S-\overline{S})\right)^2} \, + \,
3\,\frac{\partial T\partial \overline{T}̣}{\left(-i(T-\overline{T})\right)^2} \, + \,
3\,\frac{\partial U\partial \overline{U}}{\left(-i(U-\overline{U})\right)^2} \ .
\ee
A scalar potential $V$ is determined by $K$ given in \eqref{Kaehler_STU} and a holomorphic superpotential $W$ which will receive a perturbative contribution from the fluxes and a non-perturbative 
one from open-string dynamics such as \emph{e.g.} gaugino condensation \cite{Achucarro:2006zf}. This reads
\be
\label{V_N=1}
V\,=\,e^{K}\left(-3\,|W|^{2}\,+\,K^{\a\bar{\b}}\,D_{\a}W\,D_{\bar{\b}}\overline{W}\right)\ ,
\ee
where $K^{\a\bar{\b}}$ is the inverse K\"ahler metric and $D$ denotes the K\"ahler-covariant derivative.

\subsection*{The set-up}

We consider the following reduction \emph{Ansatz}
\be
\label{IIA_red_ansatz}
ds^{2}_{10} \, = \, \tau^{-2} \, ds^{2}_{4} \, + \, \rho \, \left(\sigma^{-3} \, M_{ab} \, dy^{a} \, dy^{b} \, + \, \sigma^{3} \, M_{ij} \, dy^{i} \, dy^{j}\right) \ ,
\ee
where the universal moduli $\tau$ and $\rho$ are defined as the following combinations \cite{Hertzberg:2007wc}
\be
\left\{\begin{array}{lclc}
\rho & = & \left(\textrm{vol}_{6}\right)^{1/3} & , \\
\tau & = & e^{-\phi} \, \sqrt{\textrm{vol}_{6}} & 
\end{array}\right.
\ee
of the internal volume and ten-dimensional dilaton, while $\sigma$ fixes the relative ratio between the volume of the three-cycle along $y^{a}$ and $y^{i}$, respectively. 
Their expressions in terms of the supergravity fields $\Phi^{\a}$ are given by\footnote{Please note that, in a type IIA language, the ten-dimensional dilaton corresponds to a combination
of $S$ and $T$, and the role of K\"ahler and complex structure moduli are interchanged w.r.t. the standard conviontions of type IIB with O$3$ and O$7$-planes.}
\be
\label{rhotausigma/STU}
\left\{
\begin{array}{lclc}
\rho & = & \textrm{Im}(U) & , \\
\tau & = & \textrm{Im}(S)^{1/4} \, \textrm{Im}(T)^{3/4} & , \\
\sigma & = & \textrm{Im}(S)^{-1/6} \, \textrm{Im}(T)^{1/6} & . 
\end{array}
\right.
\ee

Furthermore, we place the following D$6$-branes as local sources, divided into 
\be
\begin{array}{lcccc}
\textrm{D}6^{||} \, : &  &  & \underbrace{\times \, \vert \, \times \, \times \, \times}_{D = 4} \,  \, \underbrace{\times \, \times \, \times}_{a} \,\underbrace{ - \, - \, -}_{i} & ,
\end{array}
\notag
\ee
and 
\be
\begin{array}{lcccccc}
\textrm{D}6^{\perp}  \, : &  &  & \underbrace{\times \, \vert \, \times \, \times \, \times}_{D = 4} \, &  \, \underbrace{\begin{array}{c}
- \, - \, \times \, \\
- \, \times \, - \, \\
\times \, - \, - \,   
\end{array}}_{a} & \underbrace{\begin{array}{c}
\times \, \times \, - \, \\
\times \, - \, \times  \, \\
- \, \times \, \times \,   
\end{array}}_{i} & ,
\end{array}
\notag
\ee
the three latter ones being identified among them in the isotropic case.

\subsection*{The perturbative superpotential}

The most general set of isotropic geometric fluxes in this duality frame is collected in table~\ref{table:fluxes}.
\begin{table}[h!]
\renewcommand{\arraystretch}{1.25}
\begin{center}
\scalebox{0.92}[0.92]{
\begin{tabular}{ | c || c | c |}
\hline
couplings & Type IIA & parameters \\
\hline
\hline
$1 $ &  $F_{aibjck}$ & $a_0 $ \\
\hline
$U$ &  $F_{aibj}$ & $-a_1$ \\
\hline
$U^{2}$ &  $F_{ai}$ & $a_2$ \\
\hline
$U^{3}$ & $F_{0}$ & $-a_3 $ \\
\hline
\hline
$S$ & $ {H}_{ijk} $ &  $-b_0$ \\
\hline
$S \, U$ &  ${{\omega}_{ij}}^{c}$ & $b_1$ \\
\hline
\hline
$T$ & $ H_{a b k} $ & $c_0$ \\
\hline
$T \, U$ &  $ {\omega_{k a}}^{j} = {\omega_{b k}}^{i} \,\,\,,\,\,\, {\omega_{b c}}^a $  & $c_1$ \\
\hline
\end{tabular}
}
\end{center}
\caption{{\it Mapping between fluxes and couplings in the superpotential in type IIA with O$6$/D$6$. The six internal directions of $T^{6}$ are split into $a=1,2,3$ and $i=4,5,6$, as described in 
\eqref{IIA_red_ansatz}.}}
\label{table:fluxes}
\end{table}
By including these fluxes, one induces the following superpotential
\be
W_{\textrm{(pert.)}} \, = \, \left(a_{0} \, - \, 3a_{1} \, U \, + \, 3a_{2} \, U^{2} \, - \, a_{3} \, U^{3}\right) \, - \, 
\left(b_{0} \, - \, 3b_{1} \, U \right) \, S \, + \, \left(c_{0} \, + \, c_{1} \, U \right) \, T \ .
\ee
The above fluxes induce a tadpole given by
\be
\label{tadpoles}
\begin{array}{lcccl}
N_{6}^{||} \, \equiv \, a_{3} \, b_{0} \, - \, 3a_{2} \, b_{1} &  & \textrm{and} & & N_{6}^{\perp} \, \equiv \, - \, a_{3} \, c_{0} \, - \, a_{2} \, c_{1}
\end{array}
\ee
for the corresponding local sources $\textrm{O}6^{||}/\textrm{D}6^{||}$ and $\textrm{O}6^{\perp}/\textrm{D}6^{\perp}$, respectively. In order to be able to interpret the underlying compactifications
as arising from pure D-brane sources, one needs $N_{6}^{||}\,>\,0$ and $N_{6}^{\perp}\,>\,0$. Please note that the set of perturbative fluxes introduced in table~\ref{table:fluxes} has been checked to
already satisfy all open and closed string Bianchi identities provided that the conditions \eqref{tadpoles} are satisfied.

\subsection*{The non-perturbative superpotential}

Given a YM theory with gauge group $\textrm{SU}(N)$, when gaugino condensation takes place, then a non-perturbative superpotential is induced which goes as $e^{-\alpha/g_{\textrm{YM}}^{2}}$, where 
$\alpha \, \equiv \, \frac{2\pi}{N}$.
The YM couplings of the gauge theories living on $\textrm{D}6^{||}$ and $\textrm{D}6^{\perp}$ are, respectively, given by
\be
\label{non-pert-S}
\begin{array}{cccccccc}
\dfrac{1}{\left(g_{\textrm{YM}}^{||}\right)^{2}} & = & \dfrac{\textrm{vol}^{||}_{3}}{g_{s}} & \sim & \dfrac{\rho^{3/2}}{g_{s}} \, \sigma^{-9/2} & \equiv & \textrm{Im}(S) & , 
\end{array}
\ee
and
\be
\label{non-pert-T}
\begin{array}{cccccccc}
\dfrac{1}{\left(g_{\textrm{YM}}^{\perp}\right)^{2}} & = & \dfrac{\textrm{vol}^{\perp}_{3}}{g_{s}} & \sim & \dfrac{\rho^{3/2}}{g_{s}} \, \sigma^{3/2} & \equiv & \textrm{Im}(T) & .
\end{array}
\ee

Motivated by this argument, we will consider the following non-perturbative superpotential
\be
\label{W_nonpert}
W_{\textrm{(non-pert.)}} \, = \, \left(Z_{1} \, + \,iZ_{2} \right) \, e^{i\a\,S} \, + \,  \left(Z_{3} \, + \,iZ_{4} \right) \, e^{i\b\,T} \ ,
\ee
where the constants $Z_{r}$, for $r=1,\,2,\,3,\,4$ are real and the positive numbers $\a$ and $\b$ are related to the rank of the corresponding gauge groups as explained above.

In section~\ref{sec:solution}, we will show how the scalar potential deriving from $W \, \equiv \, W_{\textrm{(pert.)}} \, + \, W_{\textrm{(non-pert.)}}$ as shown in \eqref{V_N=1} contains examples of stable dS
solutions.

\section{Suppressed non-perturbative corrections}
\label{sec:S&S}

In this section, we will show how to produce dS critical points by adding non-perturbative corrections to a set of perturbative background fluxes admitting a stable dS solution up to a run-away
in, say $\textrm{Im}(T)$. An example of such a background is given in table~\ref{example}, which also happens to have positive $N_{6}^{||}$ and $N_{6}^{\perp}$ as defined in \eqref{tadpoles}.
\begin{table}[h!]
\renewcommand{\arraystretch}{1.25}
\begin{center}
\scalebox{0.90}[0.90]{
\begin{tabular}{ | c || c |}
\hline
Fluxes  &  Example \\
\hline 
\hline
$\begin{array}{cc} a_{0} & a_{1} \\ a_{2} & a_{3} \\ b_{0} & b_{1} \\ c_{0} & c_{1}\end{array}$ & $\begin{array}{cc} 0 & 0 \\ \frac{1}{2} \left(3-\sqrt{29}\right) & 
-\sqrt{\frac{3}{14} \left(33+5 \sqrt{29}\right)} \\ \frac{1}{28} \left(3 \sqrt{42 \left(33+5 \sqrt{29}\right)}-\sqrt{1218 \left(33+5 \sqrt{29}\right)}\right) & 1 \\ 
0 & 1 \end{array}$  \\
\hline
\end{tabular}
}
\end{center}
\caption{{\it Example of a flux background having a dS critical point at $S\,=\,U\,=\,i$, $\textrm{Re}(T)\,=\,0$ with a run-away behaviour in $\textrm{Im}(T)$. The value of $V_{0}$ is $\frac{3}{16} 
\left(6+\sqrt{29}\right)$, whereas the tadpoles read $N_{6}^{||}\,=\,\frac{3}{7} \left(1+8 \sqrt{29}\right)$ and $N_{6}^{\perp}\,=\,\frac{1}{2} \left(\sqrt{29}-3\right)$.}}
\label{example}
\end{table}

If one now adds a $T$-dependent non-perturbative contribution to the superpotential (like in equation \eqref{W_nonpert} with $Z_{1} = Z_{2} = 0$) and takes the limit $\textrm{Im}(T) \, \equiv \, \lambda \, \gg \, 1$, 
one schematically finds (up to subleading contributions)
\be
\begin{array}{lclc}
V & = & V_{(\textrm{pert.})} \, + \, \mathcal{O}\left(Z \,e^{-\beta\lambda}\right) & ,
\end{array}
\ee
where $Z \, \equiv \, \sqrt{Z_{3}^{2}+Z_{4}^{2}}$, and
\be
\begin{array}{lclc}
\partial_{\phi_{i}}V & = & \partial_{\phi_{i}}V_{(\textrm{pert.})} \, + \, \mathcal{O}\left(Z \, e^{-\beta\lambda}\right) & ,
\end{array}
\ee
for any field $\phi_{i}$ other than $\textrm{Im}(T)$. The eom for this direction will, instead, receive important non-perturbative contribution of the size
\be
\label{eom_T}
\begin{array}{lclc}
\partial_{\textrm{Im}(T)}V & = & \partial_{\textrm{Im}(T)}V_{(\textrm{pert.})} \, + \, \mathcal{O}\left(Z \, \beta \lambda \, e^{-\beta\lambda}\right) & ,
\end{array}
\ee
at least in a regime in which there is some competition between exponetial suppression and the power-law growth in the second term above.

In general, one can solve \eqref{eom_T} w.r.t. $Z_{3}$ and $Z_{4}$ and make sure that $\textrm{Im}(T)$ sits at a minimum where also the masses of the other moduli receive small contributions. The only problem 
is that the off-diagonal mixing in the mass matrix, \emph{i.e.} ${\left(m^{2}\right)_{\textrm{Im}(T)}}^{\phi_{i}}$, will, generically, receive contributions of the same size, thus leading to tachyons
upon diagonalilsation. This feature represents the main difference w.r.t. ref.~\cite{Kachru:2003aw}, where such tachyons are avoided by means of a separation between complex structure and K\"ahler moduli
realised by the uplifting anti-brane potential which is manifestly independent of all the complex structure moduli. In our case, such uplifting is provided by the supergravity F-terms, which non-trivially 
mix the two sectors.
The aforementioned problem arising in this context can therefore only be solved by explicitly looking for special fine-tuned situations where such a separation is artificially introduced. 
It still remains to be seen whether examples of this type can actually exist in type IIA.

\section{The stable dS examples}
\label{sec:solution}

In this section we will systematically search for stable dS solutions arising from IIA compactifications with combined effect of perturbative fluxes and non-perturbative effects.
First of all, we exploit the homogeneity of our scalar manifold to identify a special point
\be
\label{origin}
\begin{array}{cccccccc}
S_{0} & = & T_{0} & = & U_{0} & = & i & ,
\end{array}
\ee
called the origin of moduli space. Looking for theories having a critical point in the origin translates the eom's for the six real scalar fields into quadratic equations \cite{Dibitetto:2011gm} in the 
superpotential parameters. This set of quadratic equations is what we are going to solve in this section.

The total number of real parameters in the superpotential 
\be
\left\{a_{0},\,a_{1},\,a_{2},\,a_{3},\,b_{0},\,b_{1},\,c_{0},\,c_{1};\,\,Z_{1},\,Z_{2},\,Z_{3},\,Z_{4}\right\} 
\notag
\ee
exactly coincides with twice
the number of real fields, \emph{i.e.} $12$. This makes it possible to apply the technique used in ref.~\cite{Danielsson:2012by} based on the following linear change of variables
\be
\label{SUSY_break}
D_{\a}W|_{\Phi_{0}}\,=\,A_{\a} \,+\, iB_{\a} \ , 
\ee
where $A_{\a}$ and $B_{\a}$ denote $6$ real constants called supersymmetry-breaking parameters. Once $6$ superpotential couplings are exchanged for the supersyemmetry-breaking parameters by solving the 
linear equations in \eqref{SUSY_break}, the $6$ remaining fluxes only appear linearly in the (orginally) quadratic field equations in the origin of moduli space. 

The resulting parameter space of solutions is therefore six-dimensional. By randomly choosing numerical values for $\left\{A_{\a} ,\, B_{\a}\right\}$, one can scan such a space of solutions looking
for interesting regions. Within $\mathcal{O}(10^{5})$ random choices, we were able to find two stable dS critical points having fully positive non-isotropic mass spectrum and the correct sign of the 
tadpoles in \eqref{tadpoles}. The deatils of the two solutions which were found are given in tables~\ref{dS_sols1} and \ref{dS_sols2}.
\begin{table}[h!]
\renewcommand{\arraystretch}{1.25}
\begin{center}
\scalebox{0.90}[0.90]{
\begin{tabular}{ | c || c | c |}
\hline
Fluxes  &  Sol.~$1$ & Sol.~$2$ \\
\hline 
\hline
$\begin{array}{cc} a_{0} & a_{1} \\ a_{2} & a_{3} \\ b_{0} & b_{1} \\ c_{0} & c_{1}\end{array}$ & $\begin{array}{cc} 0.378482 & -0.335967 \\ -0.120278 & -0.135393 \\ -0.273515 & 0.029837 \\ 
-0.019665 & 0.027261 \end{array}$  &  
$\begin{array}{cc} -0.445792 & -0.253070 \\ 0.072296 & 0.093816 \\ 0.226982 & 0.097918 \\ -0.048988 & 0.021274 \end{array}$\\
\hline
$\begin{array}{cc} Z_{1} & Z_{2} \\ Z_{3} & Z_{4} \end{array}$ & $\begin{array}{cc} -0.385558 & -1.00688 \\ 3.24742 & 3.23091 \end{array}$  & 
$\begin{array}{cc} 0.973554 & -0.114369 \\ -0.065455 & -0.416440 \end{array}$ \\
\hline
\end{tabular}
}
\end{center}
\caption{{\it The two stable dS extrema found by random search. The first part of the table shows the values of the $8$ perturbative fluxes, 
whereas in the second part we give the explicit values of the $4$ parameters driving the non-perturbative effects. Please note that the location of the critical points is chosen to be the origin of 
moduli space and the parameters $\a$ and $\b$ are for simplicity chosen equal to $1$.}}
\label{dS_sols1}
\end{table}
\begin{table}[h!]
\renewcommand{\arraystretch}{1.25}
\begin{center}
\scalebox{0.90}[0.90]{
\begin{tabular}{ | c || c | c |}
\hline
  &  Sol.~$1$ & Sol.~$2$ \\
\hline 
\hline
$V_{0}\,\equiv\,V(\Phi_{0})$ & $3.55364\,\times\,10^{-4}$ & $2.25745\,\times\,10^{-6}$ \\
\hline
$m_{3/2}^{2}\,\equiv\,\left|W(\Phi_{0})\right|^{2}$ & $0.425233$ & $0.551397$ \\
\hline
$\begin{array}{ccc} N_{6}^{||} & & N_{6}^{\perp} \end{array}$ & $\begin{array}{cc} 0.047798 & 6.16361\,\times\,10^{-4} \end{array}$  &  $\begin{array}{cc} 5.74369\,\times\,10^{-5} & 3.05783\,\times\,10^{-3}\end{array}$ \\
\hline
\begin{tabular}{c} Normalised \\[-3mm] masses \\ $(m^{2}/V_{0})$ \end{tabular}
& $\begin{array}{cc} 347.232 & 168.749 \\ 35.2127 \,\,(2\times) & 30.3856 \\ 20.4299 \,\,(2\times) & 11.7289 \\ 7.12074 & 6.69231 \,\,(2\times) \\ 1.53982 & 1.28715 \,\,(2\times) \end{array}$ & 
$\begin{array}{cc} 19790.3 & 10976.4 \\ 7439.16 & 4889.19 \,\,(2\times) \\ 3289.23 \,\,(2\times) & 3212.11 \\ 945.310 \,\,(2\times) & 708.254 \\ 125.691 & 49.0627 \,\,(2\times) \end{array}$ \\
\hline
\end{tabular}
}
\end{center}
\caption{{\it The physical quantities characterising the two found stable dS. The first row shows the values of the cosmological constant, the second one the gravitino mass, the third row the values of 
the tadpoles for the local sources, and the fourth one the full non-isotropic mass spectra normalised to the cosmological constant. Please note that isotropic backgrounds always produce six 
non-degenerate masses in the spectrum, whereas the remaining eight directions are grouped into four pairs of double eigenvalues.}}
\label{dS_sols2}
\end{table}

\section{Comments on scale separation}
\label{sec:discussion}

In ref.~\cite{Blaback:2013fca}, it was already noticed that scale separation can in principle be achieved within geometric type IIA compactifications at the price of tuning down metric flux in the large 
volume limit, whereas all the other fluxes become large and hence insensitive to flux quantisation. 
However, looking more carefully, it was also noticed that compactifications of this type with O$6$-planes rather than D$6$-branes all present the same problem when analysing tadpole cancellation in the 
limit of large fluxes. This is related to the fact that the orientifold charge is fixed by the topology of the internal manifold and is typically $-\mathcal{O}(1)$. 
Introducing non-perturbative effects can lead to dS solutions without O-planes, thus overcoming the aforementioned issue. In addition, it is worth mentioning that the same logic of \cite{Blaback:2013fca} 
can be applied here since the rescaling of fluxes and scalars used to move away from \eqref{origin} can be generalised to a symmetry of the full scalar potential including non-perturbative effects.

In this paper we have shown how to make use of non-perturbative effects in order to find stable dS solutions with pure D-brane sources. In this context, one can perform a similar analysis without encountering
the problems related to tadpole cancellation. Moreover, one finds that the large-volume scale $N$ first introduced in ref.~\cite{DeWolfe:2005uu} coincides\footnote{This can be seen easily by looking at how $\a$ and $\b$
(defined as $\frac{2\pi}{N}$) appear in \eqref{W_nonpert}.} with the rank of the $\textrm{SU}(N)$ gauge groups realised on the different types of D$6$-branes (which are naturally identified when only 
restricting to the universal moduli).

On the other hand, though, this interpretation of the scale $N$ coming from the gauge theory living on the D$6$-branes brings a new requirement into the game, \emph{i.e.} the tadpoles introduced in 
\eqref{tadpoles} and defining the number of D$6$-branes should scale faster than or equally fast as $N$ itself. The former case corresponds to a situation in which a given subsector of the gauge theory
participates in the condensation process, whereas the latter one represents the critical situation where all the dof's are involved in such a process.

In particular, if one focuses on this special case, by scaling the universal moduli as $\rho \, \sim \, N^{1/2}$ and $\tau \, \sim \, N$ (which translates into $S \, \sim \, T \, \sim \, N$ and 
$U \, \sim \, N^{1/2}$), the perturbative flux quanta as
\be
\begin{array}{lclclclclclc}
f_{0} & \sim & N^{1/4} & , &  f_{2} & \sim & N^{3/4} & , & f_{4} & \sim & N^{5/4} & , \\[2mm]
f_{6} & \sim & N^{7/4} & , &  h_{3} & \sim & N^{3/4} & , & \omega & \sim & N^{1/4} & , 
\end{array}
\notag
\ee
and the non-perturbative parameters as $\a,\,\b \, \sim \, N^{-1}$ and $Z_{r} \, \sim \, N^{1/4}$, for $r=1,\,2,\,3,\,4$, one achieves $g_{s} \, \sim \, N^{-1/4}$ and\footnote{These powers for the 
scaling behaviours of the different physical scales in the problem refer to the Einstein frame, where $M_{\textrm{Pl}}$ is kept constant equal to $1$.}
\be
\label{scalings}
\begin{array}{cccccc}
\underbrace{\quad R\quad }_{\sim \, N^{5/4}} & \gg & \underbrace{\quad \ell_{s}\quad }_{\sim \, N^{1}} & \gg & 
\underbrace{\quad \ell_{\textrm{Pl}}\quad }_{\sim \, N^{0}} & ,
\end{array}
\ee
which implies a perturbative regime, and a separation among the KK, string and Planck scale, in the large $N$ limit. In such a regime, the tadpoles in \eqref{tadpoles} scale exactly as $N$, as we wanted.

However, in this context it turns out to be impossible to get a genuine separation of the Hubble scale based on power-counting. Instead, one finds
\be
\label{scaling_H}
\begin{array}{lclc}
H^{-1} & \sim & N^{1} & ,
\end{array}
\ee
which is of the same order of the string scale itself.

Nevertheless, beyond what suggested by power-counting arguments, one could still make use of some extra fine-tuning in order to force the Hubble scale to grow faster than what indicated in 
\eqref{scaling_H} (just corresponding to $m_{3/2}^{-1}$) and reach values of the cosmological constant which might be phenomenologically more appealing. 
Note that this possibility is somehow suggested by the distribution of such stable dS solutions along a narrow band close to the Minkowski line, in a similar way as previously found in refs~
\cite{Danielsson:2012by, Blaback:2013ht} in the context of non-geometric flux compactifications. Here, the typical values of the cosmological constant are at least four or five orders of magnitude smaller
 than $m_{3/2}^{2}$ (see table~\ref{dS_sols2}) and the previous argument may be seen as an indication that there might be an unlimited amount of fine-tuning available within the stable dS region bringing 
down to the Minkowski line. 

Another possible issue related to the general scaling properties of this class of models, as one can see from \eqref{scalings}, is that metric flux grows large even when normalised to the KK scale, \emph{i.e.}
$\omega \, R$ grows with $N$ in this limit. However, the possibile existence of highly fine-tuned internal manifolds with such a curvature cannot be ruled out.

\section{Conclusions}

Summarising, in this paper we have analysed the possibility of using a combination of perturbative fluxes and non-perturbative effects within (geometric) type IIA compactifications with the
aim of finding stable dS solutions without anti-branes and orientifold planes. We have, firstly, been able to argue that doing without the aforementioned ingredients is actually possible as far as the existence
of dS critical points is concerned. This can be combined with the requirement that the cosmological constant receives negligible contributions from non-perturbative effects in the large volume
limit.

Nevertheless, when moving to the analysis of the mass matrix, we have observed a crucial difference when using F-terms for uplifiting instead of an anti-brane potential. Due to a big mixing between the
sector of those moduli which get fixed by means of non-perturbative corrections and the others, tachyons usually arise. We leave the discussion about the possible existence of special fine-tuned tachyon-free
examples in such a regime for future work. 

Secondly, by systematically scanning the parameter space of solutions, we have found two examples of (even non-isotropically) stable dS backgrounds exhibiting purely D-brane charges for all allowed local 
sources. The only ingredients that remain, however, to some extent controversial, are the nonperturbative terms. This is due to the fact that in these examples the non-perturbative contributions to the
energy are important and hence their reliability as honest classical solutions is completely unclear.  

In particular, one related issue is our assumption that the prefactors of the non-perturbative terms inside the superpotential are independent of the other moduli. Since there is no separation between those
moduli entering exponentially $W_{(\textrm{non-pert.})}$ and the other ones, this assumption might not be correct in general. Thus, it would be interesting to study a case where such moduli dependence of 
the prefactors can be computed in order to check how this would affect the analysis done here.

Furthermore, even though we do not need to introduce any anti-branes, instabilities similar to those observed in, \emph{e.g.}, \cite{Bena:2009xk} could still be relevant. 
To find out, one would need to study the localised brane solutions in ten dimensions, rather than the smeared version that we implicitly use for our supergravity analysis. We do not address this issue in 
the present work. 

From the viewpoint of the corresponding lower-dimensional supergravity description, a possible direction in which to extend this work would be to check the mass matrix for the orientifold-odd scalars, 
which have been consistently truncated away in the present work. The consistency of the truncation only implies the existence of the same critical points once those new excitations are made dynamical,
but their stability is not guaranteed. The explicit computation of the full mass matrix has been carried out in ref.~\cite{Dibitetto:2012ia} in case of total absence of sources (\emph{i.e.} not only O-planes, but also D-branes),
 where the orientifold-odd sector was only found to be tachyon-free in some cases. So, generically one should expect to have a similar situation here.

Finally, we discuss some physical features of the stable solutions found in this way and still, we observe some interesting properties such as the possibility of achieving separation between Planck, string
and KK scale without suffering from those problems related to tadpole cancellation conditions which are typical of backgrounds with O-planes. The Hubble scale, instead, can only be separated by means of an 
extra fine-tuning.

%
%

\section*{Acknowledgments}

We would like to thank Johan Bl\r{a}b\"ack, Thomas Van Riet for stimulating discussions. We would also like to thank Thomas Van Riet for valuable comments on a draft version of this manuscript. 
The work of the authors is supported by the Swedish Research Council (VR), and the G\"oran Gustafsson Foundation.


%
%

%
%

\small

\bibliography{DD2}

\providecommand{\href}[2]{#2}\begingroup\raggedright\begin{thebibliography}{10}

\bibitem{Kachru:2003aw}
S.~Kachru, R.~Kallosh, A.~D. Linde, and S.~P. Trivedi, ``{De Sitter vacua in
  string theory},'' \href{http://dx.doi.org/10.1103/PhysRevD.68.046005}{{\em
  Phys.Rev.} {\bf D68} (2003)  046005},
\href{http://arxiv.org/abs/hep-th/0301240}{{\tt arXiv:hep-th/0301240
  [hep-th]}}.

\bibitem{Bena:2009xk}
I.~Bena, M.~Grana, and N.~Halmagyi, ``{On the Existence of Meta-stable Vacua in
  Klebanov-Strassler},'' \href{http://dx.doi.org/10.1007/JHEP09(2010)087}{{\em
  JHEP} {\bf 1009} (2010)  087},
\href{http://arxiv.org/abs/0912.3519}{{\tt arXiv:0912.3519 [hep-th]}}.

J.~Bl{\aa}b{\"a}ck, U.~H. Danielsson, D.~Junghans, T.~Van~Riet, T.~Wrase, {\em et al.},
  ``{The problematic backreaction of SUSY-breaking branes},''
  \href{http://dx.doi.org/10.1007/JHEP08(2011)105}{{\em JHEP} {\bf 1108} (2011)
   105},
\href{http://arxiv.org/abs/1105.4879}{{\tt arXiv:1105.4879 [hep-th]}}.

I.~Bena, D.~Junghans, S.~Kuperstein, T.~Van~Riet, T.~Wrase, {\em et al.},
  ``{Persistent anti-brane singularities},''
  \href{http://dx.doi.org/10.1007/JHEP10(2012)078}{{\em JHEP} {\bf 1210} (2012)
   078},
\href{http://arxiv.org/abs/1205.1798}{{\tt arXiv:1205.1798 [hep-th]}}.

I.~Bena, M.~Grana, S.~Kuperstein, and S.~Massai, ``{Anti-D3's - Singular to the
  Bitter End},'' \href{http://dx.doi.org/10.1103/PhysRevD.87.106010}{{\em
  Phys.Rev.} {\bf D87} (2013)  106010},
\href{http://arxiv.org/abs/1206.6369}{{\tt arXiv:1206.6369 [hep-th]}}.

I.~Bena, G.~Giecold, M.~Grana, N.~Halmagyi, and S.~Massai, ``{The backreaction
  of anti-D3 branes on the Klebanov-Strassler geometry},''
  \href{http://dx.doi.org/10.1007/JHEP06(2013)060}{{\em JHEP} {\bf 1306} (2013)
   060},
\href{http://arxiv.org/abs/1106.6165}{{\tt arXiv:1106.6165 [hep-th]}}.

J.~Bl{\aa}b{\"a}ck, U.~H. Danielsson, D.~Junghans, T.~Van~Riet, T.~Wrase, {\em
  et al.}, ``{(Anti-)Brane backreaction beyond perturbation theory},''
  \href{http://dx.doi.org/10.1007/JHEP02(2012)025}{{\em JHEP} {\bf 1202} (2012)
   025},
\href{http://arxiv.org/abs/1111.2605}{{\tt arXiv:1111.2605 [hep-th]}}.

J.~Bl{\aa}b{\"a}ck, U.~H. Danielsson, and T.~Van~Riet, ``{Resolving anti-brane
  singularities through time-dependence},''
  \href{http://dx.doi.org/10.1007/JHEP02(2013)061}{{\em JHEP} {\bf 1302} (2013)
   061},
\href{http://arxiv.org/abs/1202.1132}{{\tt arXiv:1202.1132 [hep-th]}}.

I.~Bena, J.~Bl{\aa}b{\"a}ck, U.~Danielsson, and T.~Van~Riet, ``{Antibranes
  don't go black},'' \href{http://dx.doi.org/10.1103/PhysRevD.87.104023}{{\em
  Phys.Rev.} {\bf D87} (2013)  104023},
\href{http://arxiv.org/abs/1301.7071}{{\tt arXiv:1301.7071 [hep-th]}}.

\bibitem{Uranga:2008nh}
A.~M. Uranga, ``{D-brane instantons and the effective field theory of flux
  compactifications},''
  \href{http://dx.doi.org/10.1088/1126-6708/2009/01/048}{{\em JHEP} {\bf 0901}
  (2009)  048},
\href{http://arxiv.org/abs/0808.2918}{{\tt arXiv:0808.2918 [hep-th]}}.

\bibitem{Barreiro:1998nd}
T.~Barreiro, B.~de~Carlos, J.~Casas, and J.~Moreno, ``{Anomalous U(1), gaugino
  condensation and supergravity},''
  \href{http://dx.doi.org/10.1016/S0370-2693(98)01402-6}{{\em Phys.Lett.} {\bf
  B445} (1998)  82--93},
\href{http://arxiv.org/abs/hep-ph/9808244}{{\tt arXiv:hep-ph/9808244
  [hep-ph]}}.

\bibitem{Saltman:2004sn}
A.~Saltman and E.~Silverstein, ``{The Scaling of the no scale potential and de
  Sitter model building},''
  \href{http://dx.doi.org/10.1088/1126-6708/2004/11/066}{{\em JHEP} {\bf 0411}
  (2004)  066},
\href{http://arxiv.org/abs/hep-th/0402135}{{\tt arXiv:hep-th/0402135
  [hep-th]}}.

\bibitem{Blaback:2013fca}
J.~Bl{\aa}b{\"a}ck, U.~Danielsson, and G.~Dibitetto, ``{Accelerated Universes
  from type IIA Compactifications},''
\href{http://arxiv.org/abs/1310.8300}{{\tt arXiv:1310.8300 [hep-th]}}.

\bibitem{Danielsson:2012by}
U.~Danielsson and G.~Dibitetto, ``{On the distribution of stable de Sitter
  vacua},'' \href{http://dx.doi.org/10.1007/JHEP03(2013)018}{{\em JHEP} {\bf
  1303} (2013)  018},
\href{http://arxiv.org/abs/1212.4984}{{\tt arXiv:1212.4984 [hep-th]}}.

\bibitem{Blaback:2013qza}
  J.~Bl{\aa}b{\"a}ck, D.~Roest and I.~Zavala,
  ``De Sitter Vacua from Non-perturbative Flux Compactifications,''
  arXiv:1312.5328 [hep-th].

\bibitem{Achucarro:2006zf}
A.~Achucarro, B.~de~Carlos, J.~Casas, and L.~Doplicher, ``{De Sitter vacua from
  uplifting D-terms in effective supergravities from realistic strings},''
  \href{http://dx.doi.org/10.1088/1126-6708/2006/06/014}{{\em JHEP} {\bf 0606}
  (2006)  014},
\href{http://arxiv.org/abs/hep-th/0601190}{{\tt arXiv:hep-th/0601190
  [hep-th]}}.

\bibitem{Hertzberg:2007wc}
M.~P. Hertzberg, S.~Kachru, W.~Taylor, and M.~Tegmark, ``{Inflationary
  Constraints on Type IIA String Theory},''
  \href{http://dx.doi.org/10.1088/1126-6708/2007/12/095}{{\em JHEP} {\bf 0712}
  (2007)  095},
\href{http://arxiv.org/abs/0711.2512}{{\tt arXiv:0711.2512 [hep-th]}}.

\bibitem{Dibitetto:2011gm}
G.~Dibitetto, A.~Guarino, and D.~Roest, ``{Charting the landscape of N=4 flux
  compactifications},'' \href{http://dx.doi.org/10.1007/JHEP03(2011)137}{{\em
  JHEP} {\bf 1103} (2011)  137},
\href{http://arxiv.org/abs/1102.0239}{{\tt arXiv:1102.0239 [hep-th]}}.

\bibitem{DeWolfe:2005uu}
O.~DeWolfe, A.~Giryavets, S.~Kachru, and W.~Taylor, ``{Type IIA moduli
  stabilization},'' \href{http://dx.doi.org/10.1088/1126-6708/2005/07/066}{{\em
  JHEP} {\bf 0507} (2005)  066},
\href{http://arxiv.org/abs/hep-th/0505160}{{\tt arXiv:hep-th/0505160
  [hep-th]}}.

\bibitem{Blaback:2013ht}
J.~Bl{\aa}b{\"a}ck, U.~Danielsson, and G.~Dibitetto, ``{Fully stable dS vacua
  from generalised fluxes},''
  \href{http://dx.doi.org/10.1007/JHEP08(2013)054}{{\em JHEP} {\bf 1308} (2013)
   054},
\href{http://arxiv.org/abs/1301.7073}{{\tt arXiv:1301.7073 [hep-th]}}.

\bibitem{Dibitetto:2012ia}
G.~Dibitetto, A.~Guarino, and D.~Roest, ``{Exceptional Flux
  Compactifications},'' \href{http://dx.doi.org/10.1007/JHEP05(2012)056}{{\em
  JHEP} {\bf 1205} (2012)  056},
\href{http://arxiv.org/abs/1202.0770}{{\tt arXiv:1202.0770 [hep-th]}}.

\end{thebibliography}\endgroup
\bibliographystyle{utphys}

\end{document}